# Einstein coefficients, cross sections, *f* values, dipole moments, and all that

Robert C. Hilborn

*Department of Physics, Amherst College, Anherst, MA 01002*

The relationships among various parameters describing the strength of optical transitions in atoms and molecules are reviewed. The application of these parameters to the description of the interaction between nearly monochromatic, directional light beams and atoms and molecules is given careful attention. Common pitfalls in relating these parameters are pointed out. This is a revised (February, 2002) version of a paper that originally appeared in Am. J. Phys. **50**, 982–986 (1982).

## I. INTRODUCTION

Several parameters are commonly used to describe the strength of atomic and molecular optical transitions. The Einstein *A* and *B* coefficients, *f* values (also called "oscillator strengths"), and transition dipole moments are all atomic and molecular parameters related to the "strength" of the transition. In many practical situations, on the other hand, it is useful to define an absorption coefficient (or for lasers, a gain coefficient) to describe the absorption (or amplification) of a beam of light passing through a medium consisting of the atoms or molecules of interest.

From a "kinetics" point of view, the absorption or scattering of radiation is described as a reaction or scattering process and the probability of absorption or scattering is given in terms of a cross section. It is the purpose of this paper to review the relationships among these descriptions of the light-matter interaction and to point out common pitfalls in using these relationships.

An examination of books[1-7] dealing with light-matter interaction shows a wide variety of expressions relating these parameters. Differences among these expressions involving factors of 2*p*, *e*$_o$, etc., of course, can be traced to differing units used in the definitions of the parameters. However, these differences are often exacerbated because many different (and often not clearly defined) measures of light "intensity" are used in the definitions of the parameters.

Further difficulties arise when these parameters are applied to some practical problems. The careful reader notes that the relationships among these parameters are almost always derived under the assumption that the atom is interacting with an isotropic, unpolarized, broad-band (wide frequency range) light field. This careful reader is then wary of applying these parameters to describe the interaction between atoms and directional, polarized, nearly monochromatic light beams.

Although this paper contains no new results, I believe that a unified discussion of the relationships among these parameters will prove to be of value to students, teachers, and researchers. For simplicity, the discussion will be limited to isolated atoms (e.g., atoms in a low-density gas or in an atomic beam) and to an isolated electric dipole transition between an upper level labeled 2, and a lower level labeled 1. I will use the word "atom" generically to mean the material system of interest. The index of refraction of the surrounding medium is assumed to be unity. SI units are used throughout the paper.

## II. ABSORPTION COEFFICIENT AND ABSORPTION CROSS SECTION

Let us begin the discussion by defining a phenomenological absorption coefficient $a(w)$. For a beam of light propagating in the $x$ direction, the absorption coefficient is defined by the expression

$$\frac{1}{i(w)} \frac{di(w)}{dx} = -a(w) ,  \quad (1)$$

where $i(w)$ is some measure of the power in the light beam at the frequency $w$. (The power is actually a time-averaged quantity, averaged over several optical cycles.) In most practical applications, the frequency dependence of the absorption and emission processes is important. In this paper, it is assumed that all of these frequency dependencies can be expressed in terms of a line shape function $g(w)$. The normalization of $g(w)$ is chosen so that

$$\int_{-\infty}^{+\infty} g(w) dw = 1 . \quad (2)$$

(Having the lower limit of the normalization integral be $-\infty$ greatly simplifies the normalization calculation. Negative frequencies have no special physical significance.) If the atoms are undergoing collisions (leading to collision broadening of the spectral line) or moving about (leading, in general, to Doppler broadening), then $g(w)$ describes the appropriate ensemble-average line shape.

Note that $g(w)$ has the dimensions of 1/angular frequency. One of the common pitfalls lies in not explicitly recognizing which independent variable is being used to describe the frequency or wavelength dependence. In this paper, angular frequency will be used except where noted.

We expect $\alpha(w)$ to be proportional to the number $n_1$ of atoms in level 1 (per unit volume) that the beam intercepts. (We assume that stimulated emission effects are negligible.[8]) The absorption cross section $s_a(w)$ is then defined by

$$a(w) = n_1 s_a(w) . \quad (3)$$

As long as the light is not too intense and the atomic density not too large, both $\alpha(\omega)$ and $\sigma_a(\omega)$ will be proportional to $g(\omega)$:

$$\alpha(\omega) = \alpha_0 g(\omega), \qquad (4)$$

$$\sigma_a(\omega) = \sigma_0 g(\omega). \qquad (5)$$

Note that $\alpha_0$ and $\sigma_0$ are the frequency-integrated absorption coefficient and cross section, respectively:

$$\alpha_0 = \int_{-\infty}^{+\infty} \alpha(\omega) d\omega, \qquad (6)$$

$$\sigma_0 = \int_{-\infty}^{+\infty} \sigma(\omega) d\omega, \qquad (7)$$

with dimensions of angular frequency/distance and angular frequency × area.

The frequency-integrated parameters are of limited practical usefulness. They can be used to describe the absorption and scattering of a beam of light whose frequency bandwidth is large compared to the width of $g(\omega)$. In most cases, however, we are interested in the behavior of light beams whose bandwidth is on the order of, or less than, the width of $g(\omega)$. In those cases, we must use $\alpha(\omega)$ and $\sigma_a(\omega)$.

## III. EINSTEIN COEFFICIENTS

In 1917 Einstein introduced $A$ and $B$ coefficients to describe spontaneous emission and induced absorption and emission. The Einstein $A$ coefficient is defined in terms of the total rate of spontaneous emission $W_{21}^s$ from an upper level 2 to a lower level 1 for a system of $N_2$ atoms in the upper level:

$$W_{21}^s = A_{21} N_2. \qquad (8)$$

If level 2 can decay only by radiative emission to level 1, then $A_{21}$ must be the reciprocal of the spontaneous radiative lifetime $t_{\text{spon}}$ of level 2:

$$A_{21} = 1/t_{\text{spon}}. \qquad (9)$$

(If level 2 can decay to several lower levels, the more general relation $1/t_{\text{spon}} = \sum_i A_{2i}$ must be used, where the sum is over all energy levels to which level 2 can decay.)

The $B$ coefficients are defined in terms of the transition rates for (induced) absorption $W_{12}^i$ and induced (or stimulated) emission $W_{21}^i$:

$$W_{12}^i = B_{12}^\omega \rho_\omega N_1 \qquad (10)$$

$$W_{21}^i = B_{21}^\omega \rho_\omega N_2, \qquad (11)$$

where $\rho_\omega$ is the energy density per unit angular frequency interval in the region containing $N_2$ atoms in the upper level and $N_1$ atoms in the lower level. $\rho_\omega$ is assumed to be constant over the frequency range of significant absorption and emission for the $1 \leftrightarrow 2$ transition. The $B$s have dimensions of volume × angular frequency/(energy × time).

Einstein showed that quite generally

$$B_{21}^\omega = \left(\pi^2 c^3 / \hbar \omega_{21}^3\right) A_{21} \qquad (12)$$

and

$$B_{12}^w = (g_2/g_1) B_{21}^w,  \qquad (13)$$

where $w_{21}$ is the resonance frequency of the transition. $g_1$ and $g_2$ are the degeneracy factors of the two levels.[9] As usual, $\hbar$ is Planck's constant divided by $2p$ and $c$ is the speed of light. Note that the $B$s, so defined, are independent of the details of the line shape $g(w)$ because $r_w$ is assumed to be spectrally flat over the region in which $g(w)$ varies significantly.

It is important to recognize that a different relation between the $A$ and $B$ coefficients is found if some other measure of the radiation energy density is used. For example, if we had used $r_f$ (energy density per unit frequency interval), then since $B_{21}^w r_w = B_{21}^f r_f$ and $r_w dw = r_f df$, we find that $B_{21}^f = B_{21}^w / 2p$. We use superscripts to distinguish the resulting $B$s.

To illustrate the differences that occur in the literature, three examples of the relationships of the Einstein $A$ and $B$ coefficients are quoted. (The subscript notation has been standardized.) From Herzberg (Ref. 5, p. 21), we find

$$B_{21}^n = A_{21}/(8p hc n_{21}^3),  \qquad \text{(Herzberg)} \quad (14)$$

where $n_{21} = w_{21}/(2pc)$ is the wavenumber of the transition. Yariv (Ref. 6, p. 164) shows that

$$B_{21}^f = c^3 A_{21}/(8p h f^3),  \qquad \text{(Yariv)} \quad (15)$$

where $f = w_{21}/2p$. (We do not put subscripts on $f$ to avoid possible confusion with so-called $f$-values, to be introduced below.) And finally, from Mihalas (Ref. 4, p. 79), we have

$$B_{21}^f = c^2 A_{21}/(2 h f^3)  \qquad \text{(Mihalas)} \quad (16)$$

One difference among the derivations of these relations is obvious: Herzberg defines $B$ in terms of radiation density per unit wavenumber interval while Yariv and Mihalas use radiation density per unit frequency interval. That difference accounts for a factor of $c^3$ (since $n = f/c$) between Herzberg and Yariv. The remaining factor of $c$ difference between Herzberg and Yariv arises because Herzberg defines $B$ in terms of irradiance (power per unit area and per unit wavenumber) incident on the atom, while Yariv uses energy per unit volume and per unit frequency interval at the location of the atom. The factor of $c/4p$ difference between Mihalas and Yariv arises because Mihalas has extracted a factor of $4p$ in the definition of his $B$ in terms of "specific intensity" (power per area and per solid angle and per frequency interval).

The lesson to be learned here is that before one can make use of the formulas for $A$ and $B$ coefficients from a particular source, one must carefully determine which measure of radiation intensity has been used. In fact, in the references cited above the same word "intensity" is used to signify three quite distinct physical quantities.

## IV. RELATIONSHIP BETWEEN THE EINSTEIN COEFFICIENTS AND THE ABSORPTION CROSS SECTION

Another pitfall lies in trying to apply the $B$ coefficients directly to the analysis of the behavior of nearly monochromatic, directed beams of light. Since most texts provide an inadequate treatment of this important application, the problem is discussed here in some detail.

To relate the $B$ coefficients to the absorption cross section (or alternatively, to the absorption coefficient), we need to define $B$ coefficients for monochromatic and for directed radiation. The careful reader will have already noted that the $B$ coefficients were defined above in terms of a broadband, isotropic radiation field. We first consider the case of a monochromatic (but still isotropic) field and define the induced absorption rate due to radiation in the angular frequency range from $\omega$ to $\omega + d\omega$ to be

$$w_{12}^{i}(\omega)d\omega = b_{12}(\omega)N_1 \rho(\omega)d\omega \quad , \tag{17}$$

where $\rho(\omega)d\omega$ is the energy per unit volume in the range $\omega$ to $\omega + d\omega$. [We recognize that $\rho(\omega)$ is the energy per unit volume and per unit angular frequency interval at $\omega$.]

First, we find the connection between $b_{12}(\omega)$ and $B_{12}^{\omega}$ defined previously by letting $\rho(\omega) = \rho_\omega$ be a constant over the frequency range near $\omega_{21}$. Then we integrate Eq. (17) over frequency to obtain

$$W_{12}^{i} = \int_{-\infty}^{+\infty} w_{12}^{i}(\omega)d\omega = \int_{-\infty}^{+\infty} b_{12}(\omega)N_1 \rho(\omega)d\omega$$
$$= \rho_\omega N_1 \int_{-\infty}^{+\infty} b_{12}(\omega)d\omega = \rho_\omega N_1 B_{12}^{\omega} \quad . \tag{18}$$

To incorporate the atomic frequency response explicitly, we may write

$$b_{12}(\omega) = B_{12}^{\omega} g(\omega) \quad . \tag{19}$$

We now turn to the case of directional radiation. In many practical situations, in particular when dealing with radiation transfer or lasers, it is useful to express the transition rates in terms of the irradiance (time-averaged power per unit area) of a directional beam of light. Classical and quantum-mechanical calculations show that for electric dipole transitions the absorption and stimulated emission rates depend only on the square of the amplitude of the electric field at the location of the atom (and of course on the polarization and frequency spectrum of the light). Hence, as long as the directional beam produces the same energy density (proportional to the electric field amplitude squared) at the location of the atom as does the isotropic field, the transition rate will be the same (taking polarization into account).

The irradiance $I$ is related to the electric field <u>amplitude</u> $E$ by

$$I = \tfrac{1}{2} c \varepsilon_o E^2 \quad , \tag{20}$$

where $\varepsilon_o$, as usual, is the permittivity of free space.

For a nearly monochromatic directional beam, the irradiance can be expressed in terms of an integral of $\rho(\omega)$, the energy density in the angular frequency interval between $\omega$ and $\omega + d\omega$:

$$I = \int c\rho(\omega)d\omega = \int i(\omega)d\omega \quad , \tag{21}$$

where $i(\omega)$ is the "spectral irradiance" (power per unit area and per unit angular frequency interval).[10] Then the absorption rate due to radiation in the angular frequency range $\omega$ to $\omega + d\omega$ is

$$w_{12}^i(\omega)d\omega = N_1 b_{12}(\omega)i(\omega)d\omega/c \tag{22}$$

We can now relate the $B_{12}^\omega$ coefficient to the absorption cross section $\sigma_a(\omega)$ by the following argument: Using Eq. (19) in Eq. (17), we find that power absorbed in the frequency range $\omega$ to $\omega + d\omega$ by $N_1$ atoms is $\hbar\omega B_{12}^\omega g(\omega)\rho(\omega)N_1 d\omega$. Suppose that $\rho(\omega)$ is due to a beam of cross-sectional area $A$. Then $-\Delta P = \hbar\omega B_{12}^\omega g(\omega) n_1 A d\omega \Delta x$ is the power lost from this beam as it propagates a distance $\Delta x$, where $n_1$ is the number of atoms per unit volume in level 1. The spectral irradiance in the beam is $i(\omega) = c\rho(\omega)$. With the help of

$$\frac{\Delta P}{A \Delta x d\omega} \to \frac{di}{dx}, \tag{23}$$

we find that

$$\frac{1}{i(\omega)}\frac{di(\omega)}{dx} = -\hbar\omega n_1 B_{12}^\omega g(\omega)/c. \tag{24}$$

Thus the following expressions relate the absorption cross section to the Einstein $B$ coefficient:

$$\sigma_a(\omega) = \hbar\omega B_{12}^\omega g(\omega)/c \tag{25}$$

$$\sigma_0 = \hbar\omega_{21} B_{12}^\omega/c. \tag{26}$$

In arriving at Eq. (26), we have assumed that $g(\omega)$ is sharply peaked at $\omega_{21}$, and hence that we may replace $\omega$ by $\omega_{21}$ when carrying out the integration over frequency.

We now consider the absorption process from the point of view of photons. Let $R_{12}$ denote the number of absorption events per unit time and per photon of frequency $\omega$. Then we may find the relationship between $\sigma_a$ and $B_{12}^\omega$ by using the standard expression relating the absorption rate per photon to the number density of absorbing atoms $n_1$ and the relative speed $c$ of the two collision partners:

$$R_{12} = n_1 c \sigma_a(\omega). \tag{27}$$

If we now multiply $R_{12}$ by the number of photons $dN_p(\omega)$ in a volume $V$ in the frequency range $\omega$ to $\omega + d\omega$ [$dN_p(\omega)$ is proportional to the energy in that frequency range.], we find, with the aid of $\rho(\omega)d\omega = dN_p(\omega)\hbar\omega/V$, that

$$R_{12} dN_p(\omega) = n_1 c \sigma_a(\omega)\rho(\omega)d\omega V/\hbar\omega$$
$$= N_1 c \sigma_a(\omega)\rho(\omega)d\omega/\hbar\omega. \tag{28}$$

If we compare this result with Eq. (22), we find the relation given in Eq. (25).

Using Eq. (12) (and $\omega_{12} = 2\pi c/\lambda_{12}$), we may write the cross sections in terms of the $A$ coefficient:

$$\sigma_a(\omega) = \tfrac{1}{4}(g_2/g_1)\lambda_{21}^2 g(\omega) A_{21}, \tag{29}$$

$$\sigma_0 = \tfrac{1}{4}(g_2/g_1)\lambda_{21}^2 A_{21}.$$

For an electric dipole ("allowed") transition for a stationary, isolated atom, $g(\omega_{21})A_{21}$ is often on the order of unity. Hence, the line center absorption cross section is on the order of $\lambda_{21}^2$. We are led to picture the "collision" between a photon and an atom as a collision between a fuzzy ball (the photon) of radius about equal to $\lambda_{21}$, and an atom that is small compared to $\lambda_{21}$.

For other multipole transitions, for example, for magnetic dipole and electric quadrupole transitions, the product $g(\omega_{21})A_{21}$ may be much smaller than unity if the upper level can decay via electric dipole transitions to other levels. Hence, the cross section will be correspondingly smaller. Obviously, for those other multipole transitions, the fuzzy ball picture of the photon is not appropriate[11].

## V. OSCILLATOR STRENGTH (*f* VALUE)

Oscillator strengths (*f* values) may be defined by comparing the emission rate or absorption rate of the atom with the emission or absorption rate of a classical, single-electron oscillator (with oscillation frequency $\omega_{21}$). We define an <u>emission</u> oscillator strength $f_{21}$ by the relation[2]

$$f_{21} = -\tfrac{1}{3} A_{21}/\gamma_{cl} , \tag{30}$$

where

$$\gamma_{cl} = e^2 \omega_{21}^2 /(6\pi\varepsilon_o mc^3) . \tag{31}$$

Note that we have used subscripts on the oscillator strength $f_{21}$ to distinguish it from the transition frequency. Here, *m* is the mass of the electron. The classical radiative decay rate of the single-electron oscillator at frequency $\omega_{21}$ is given by $\gamma_{cl}$. An <u>absorption</u> oscillator strength $f_{12}$ is then defined by

$$g_1 f_{12} \equiv -g_2 f_{21} \equiv gf . \tag{32}$$

The *f*s have been defined so that if (a) $g_2 = 3$ (that is, the angular momentum quantum number $J_2$ of the upper level is equal to unity), (b) $g_1 = 1$ (that is, $J_1 = 0$), and (c) the Einstein *A* coefficient is equal to the classical decay rate ($A_{21} = \gamma_{cl}$), then the resulting absorption *f* value $f_{12}$ is equal to unity and $f_{21} = -1/3$. Tables of *gf* values for many atomic transitions have been compiled.[12-14] We may now relate the absorption oscillator strength to the *A* value:

$$f_{12} = (g_2/g_1) 2\pi\varepsilon_o mc^3 A_{21} /(\omega_{21}^2 e^2) . \tag{33}$$

Alternatively, we may define the absorption oscillator strength by comparing the absorption cross section of a classical oscillator with that determined by the *B* coefficients. For a stationary, classical oscillator, the absorption cross section is[2]

$$\sigma_{ac}(\omega) = \frac{\gamma_{cl}/(2\pi)}{(\omega-\omega_o)^2 + (\gamma_{cl}/2)^2} \frac{\pi e^2}{2\varepsilon_o mc} . \tag{34}$$

Note that $\gamma_{cl}$ is the full-width-at-half-maximum of the absorption curve. With the aid of Eq. (25), we define the absorption oscillator strength $f_{12}$ by the expression

$$f_{12} = S_0 / S_{0c} = \hbar w_{21} B_{12}^w / (c S_{0c}) , \quad (35)$$

where

$$S_{0c} = \int_{-\infty}^{+\infty} S_{ac}(w) dw . \quad (36)$$

Inserting the result stated in Eq. (34) into Eq. (35), we find with the aid of Eqs. (12) and (13) the result given in Eq. (33).

## VI. TRANSITION DIPOLE MOMENT AND LINE STRENGTH

From a quantum electrodynamics treatment of spontaneous emission, it may be shown that[2,7]

$$A_{21} = \frac{2e^2 w_{21}^3}{3 e_o hc^3} \sum_{m_1} |\langle 1 m_1 | \vec{r} | 2 m_2 \rangle|^2 , \quad (37)$$

for a transition from sublevel $m_2$ of the upper level 2 to all possible $m_1$ sublevels of the lower level 1. (The usual approximations leading to the electric dipole form of the transition moment have been made. $\vec{r}$ stands for the sum of the electrons' position vectors.) Note that

$$\sum_{m_1} |\langle 1 m_1 | \vec{r} | 2 m_2 \rangle|^2$$

must be independent of $m_2$. Otherwise, the different $m_2$ levels would have different lifetimes, which is not possible in an isotropic environment.

If we have a nondegenerate two-state atom, there is only one $m_1$ and one $m_2$ and we may unambiguously define the square of the transition dipole moment $m_{21}$ by the relation

$$e^2 |\langle 1 m_1 | \vec{r} | 2 m_2 \rangle|^2 \equiv e^2 r_{21}^2 \equiv m_{21}^2 . \quad (38)$$

If the lower level is degenerate, we then define

$$e^2 \sum_{m_1} |\langle 1 m_1 | \vec{r} | 2 m_2 \rangle|^2 \equiv e^2 r_{21}^2 \equiv m_{21}^2 \quad (39)$$

as the square of the transition dipole moment. Note again that this moment is independent of $m_2$. In either case we find

$$m_{21}^2 = A_{21} \frac{3 e_o hc^3}{2 w_{21}^3} . \quad (40)$$

The "line strength" $S_{21}$ of a transition is defined by the following expression, which is symmetrical in the upper and lower state labels:

$$S_{21} = S_{12} \equiv \sum_{m_1, m_2} |\langle 1 m_1 | \vec{r} | 2 m_2 \rangle|^2$$
$$= g_2 e^2 r_{21}^2 = g_2 m_{21}^2 . \quad (41)$$

Hence, $S_{21}$ is related to $A_{21}$ by

$$S_{21} = \frac{3e_o hc^3}{2w_{21}^3} g_2 A_{21} . \tag{42}$$

The line strength $S_{12}$ is the same as the absolute-value-squared of the "reduced matrix element" of the electrons' position operators. In the quantum theory of angular momentum[15], the reduced matrix element $\langle 1\|r\|2\rangle$ of a vector operator is defined as

$$\langle 1m_1 | r_q | 2m_2 \rangle = (-1)^{J_1-m_1} \begin{pmatrix} J_1 & J_2 & 1 \\ m_1 & m_2 & q \end{pmatrix} \langle 1\|r\|2\rangle , \tag{43}$$

where $q = 0, \pm 1$ and $J_i$ is the angular momentum quantum number for the $i$th level. The factor in parentheses is the Wigner 3-j symbol. The relation with line strength is

$$S_{12} = |\langle 1\|r\|2\rangle|^2 . \tag{44}$$

## VII. RABI FREQUENCY

In many applications involving the interaction of laser radiation with atomic and molecular systems, coherent effects are important. For example, the population difference in a two-level system driven by a beam of coherent radiation oscillates in time[16-18] with an angular frequency called the "Rabi frequency." The more intense the light source, the more rapidly the population difference oscillates. The Rabi frequency for on-resonance excitation (when the frequency $w$ of exciting light equals the resonance frequency $w_{21}$) can be expressed in terms of the electric field <u>amplitude</u> of the linearly polarized light field and the transition dipole moment $m_{21}$ by the following expression[17,18]:

$$w_R = m_{21} E / \hbar . \tag{45}$$

If the upper level and lower level are degenerate, there may be several Rabi frequencies, one for each value of $\langle 1m_1 | r | 2m_2 \rangle$. In that case, the dynamical behavior of the system may be quite complicated.[18]

Some confusion in the definition of the Rabi frequency occurs because many authors make use of the so-called rotating-wave approximation, in which $E\cos(wt)$ is replaced by $\mathrm{E}\left[e^{iwt} + e^{-iwt}\right]$. Terms with a time dependence $e^{2iwt}$ are then dropped from the equations of motion. In terms of $\mathrm{E}$, the Rabi frequency is

$$w_R = m_{21} 2\mathrm{E} / \hbar . \tag{46}$$

Unfortunately, it is not always obvious which form of the electric field amplitude has been adopted. Some authors use the root-mean-square electric field $E_{rms} = E/\sqrt{2}$.

## VIII. NUMERICAL EXAMPLE AND COMMENTS

It is instructive to evaluate the parameters described above for a particular atomic transition. We consider the calcium resonance transition at $l_{21}$(air) = 422.7 nm ($w_{21} = 4.5 \times 10^{15}$ rad/s). The upper level (4 $^1$P) has a radiative lifetime of 4.5 ns. The lower level is 4 $^1$S. For our purposes, we may ignore the other possible decay routes, for example, from 4 $^1$P to 3 $^1$D. For this transition, $A_{21} = 1/t_{\text{spon}} = 2.2 \times 10^8$ s$^{-1}$. The degeneracy factors are $g_2 = 3$ and $g_1 = 1$. We now evaluate the parameters to two digit precision:

$$B_{21}^w = 6.1 \times 10^{21} \text{ m}^3 \text{ (rad/s)/(J s)},$$
$$B_{12}^w = 3B_{21}^w = 1.8 \times 10^{22} \text{ m}^3 \text{ (rad/s)/(J s)},$$
$$\sigma_0 = 2.9 \times 10^{-5} \text{ m}^2 \text{ (rad/s)},$$
$$\sigma_a(\omega = \omega_{21}) = 8.5 \times 10^{-14} \text{ m}^2 = 8.5 \times 10^6 \text{ Å}^2,$$
$$f_{12} = 1.7,$$
$$\mu_{12} = 2.4 \times 10^{-29} \text{ C m} = 1.5(\text{electron charge}) \times 1 \text{ Å}.$$

In evaluating $\sigma_a(\omega_{21})$ from Eq. (25), we have assumed that
$$g(\omega_{21}) = 2/(A_{21}\pi) \tag{47}$$
i.e., that the line shape function is Lorentzian with a full-width-at-half-maximum equal to $A_{21}$.

The experimentally determined value[20] of $f_{12}$ is $1.71 \pm 0.20$. The close agreement between the measured value and the value calculated neglecting branching to the 3 $^1$D level confirms the assumption of the smallness of that branching ratio.

Note that the line center absorption cross section (as mentioned previously) turns out to be close to $\lambda^2$ ($\lambda^2 = 1.8 \times 10^7$ Å$^2$). The transition dipole moment agrees with the expected order-of-magnitude estimate of an electron charge times a typical atomic distance.

Finally, we calculate the electric field amplitude (and corresponding irradiance), which makes the Rabi frequency equal to the natural radiative decay rate. It can be shown[6] that this is roughly the field that is required to "saturate" the atomic transition. Using the values given above we find $E \cong 10^3$ V/m and $I \cong 1.3 \times 10^3$ W/m$^2$.

All of the results developed in this paper for atoms can be immediately applied to the individual rotational transitions of molecules if the appropriate degeneracy (statistical weight) factors are used.[21,22] The definitions of transition dipole moment and line strength used here are the same as those recently recommended for the description of diatomic molecular spectra.[22]

For the convenience of the reader, Table I lists the conversion factors which relate the parameters discussed above.

**ACKNOWLEDGMENTS**

The hospitality and support of D.O. Harris and the Quantum Institute, University of California at Santa Barbara, where this work was begun, are gratefully acknowledged. I thank Mitch Trkula, Tom Hoffman, Neal Hartsough, and Ted Norris for useful comments on the manuscript. Supported in part by Research Corporation and National Science Foundation grant CHE75-23621.

| | $A_{21}$ | | | | | | |
|---|---|---|---|---|---|---|---|
| $A_{21}$ | 1 | $B_{12}^{w}$ | | | | | |
| $B_{12}^{w}$ | $\frac{g_2}{g_1}\frac{\pi^2 c^3}{\hbar w_{21}^3}$ | 1 | $B_{12}^{f}$ | | | | |
| $B_{12}^{f}$ | $\frac{g_2}{g_1}\frac{c^3}{8\pi hf^3}$ | $\frac{1}{2\pi}$ | 1 | $\sigma_0$ | | | |
| $\sigma_0$ | $\frac{g_2}{4g_1}\lambda_{21}^2$ | $\frac{\hbar w_{21}}{c}$ | $\frac{hw_{21}}{c}$ | 1 | $f_{12}$ | | |
| $f_{12}$ | $\frac{g_2}{g_1}\frac{2\pi\epsilon_o mc^3}{w_{21}^2 e^2}$ | $\frac{2\epsilon_o m\hbar w_{21}}{\pi e^2}$ | $\frac{4\epsilon_o m\hbar w_{21}}{e^2}$ | $\frac{2\epsilon_o mc}{\pi e^2}$ | 1 | $\mu_{21}^2$ | |
| $\mu_{21}^2$ | $\frac{3\epsilon_o hc^3}{2w_{21}^3}$ | $3\frac{g_1}{g_2}\frac{\epsilon_o \hbar^2}{\pi}$ | $6\frac{g_1}{g_2}\epsilon_o \hbar^2$ | $3\frac{g_1}{g_2}\frac{\epsilon_o \hbar c}{\pi w_{21}}$ | $\frac{3g_1}{2g_2}\frac{\hbar e^2}{mw_{21}}$ | 1 | $S_{21}$ |
| $S_{21}$ | $g_2\frac{3\epsilon_o hc^3}{2w_{21}^3}$ | $3\frac{g_1\epsilon_o \hbar^2}{\pi}$ | $6g_1\epsilon_o \hbar^2$ | $3\frac{g_1\epsilon_o \hbar c}{\pi w_{21}}$ | $\frac{3g_1\hbar e^2}{2w_{21}m}$ | $g_2$ | 1 |

**Table I.** Row label = entry × head of column entry, e.g. $\sigma_0 = \frac{1}{4}(g_2/g_1)\lambda_{21}^2 A_{21}$. Useful relations: $B_{12} = (g_2/g_1)B_{21}$, $f_{12} = -(g_2/g_1)f_{21}$, $m$ = mass of the electron, $e$ = charge of the electron, and $f = w_{21}/2\pi$.